# Inconsistencies in the Electronic Properties of Phosphorene Nanotubes: New Insights from Large-Scale DFT Calculations

*Sarah I. Allec and Bryan M. Wong\**

Department of Chemical & Environmental Engineering and Materials Science & Engineering Program

University of California-Riverside, Riverside, CA 92521, USA

*Corresponding author. E-mail: bryan.wong@ucr.edu. Web: http://www.bmwong-group.com

**Abstract**

Contrary to recent reports, we show that the electronic properties of phosphorene nanotubes are surprisingly rich and much more complex than previously assumed. We find that all phosphorene nanotubes exhibit an intricate direct-to-indirect bandgap transition as the nanotube diameter decreases – *a unique property not identified in any prior studies* (which claimed either direct or indirect band gaps only) that we uncover with large-scale DFT calculations. We address these previous inconsistencies by detailed analyses of orbital interactions, which reveal that the strain associated with decreasing the nanotube diameter causes a transition from a direct to an indirect band gap for *all* of the phosphorene nanotubes. We show that our findings are completely general, and extensive calculations across several exchange-correlation functionals verify our conclusions. Most importantly, our results and analyses resolve a long-standing question on the electronic properties of phosphorus nanotubes and brings closure to previously conflicting findings in these unique nanostructures.

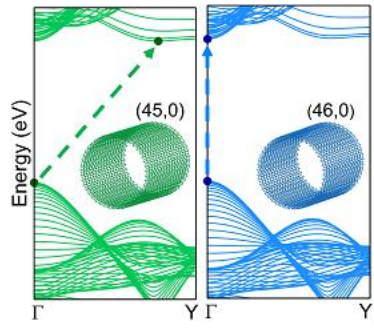

Table of Contents Graphic

Since its isolation by Novoselov *et al.* in 2004,[1] graphene has become one of the most promising materials during the last decade, particularly for electronics. However, the lack of an intrinsic band gap in graphene inhibits its use in modern electronic devices, sparking a search for other two-dimensional (2D) materials with desirable properties. To date, one of the most promising candidates is phosphorene, a single monolayer of black phosphorus arranged in a puckered honeycomb structure (Figure 1a-b). With a direct bandgap of 1-2 eV,[2-4] a carrier mobility up to 1000 $cm^2/V·s$,[5] and an on/off ratio of up to 10,000,[2, 6] phosphorene is ideal for replacing silicon in next-generation nanoelectronics. Due to the $sp^3$-hybridization in the lattice and the presence of a tunable bandgap,[7] phosphorene is also expected to exhibit other superior mechanical, electronic, and optical properties that are tunable via strain,[8-11] chemical modification,[12-14] electric gating,[6, 15-18] and nanostructuring.[19-23]

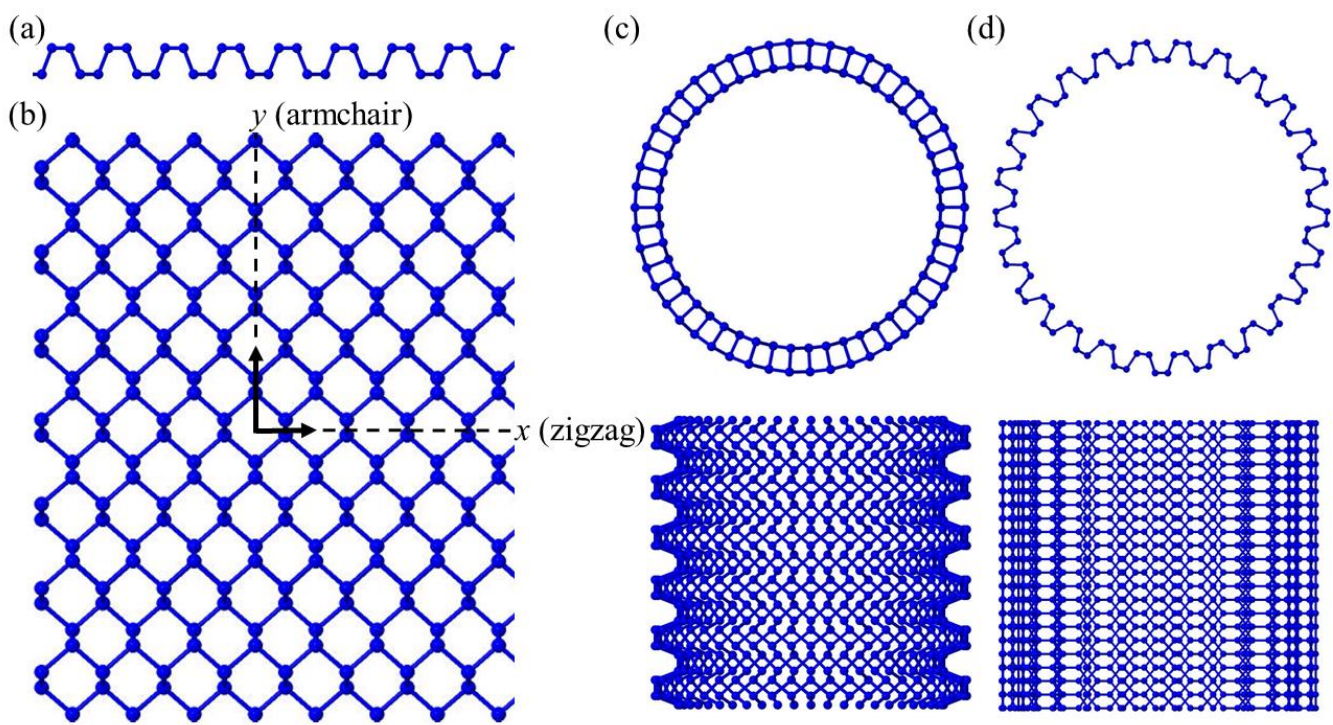

**Figure 1.** Optimized structures of phosphorene: (a) Side view and (b) Top view. Optimized structures of phosphorene nanotubes: (c) Zigzag (z-PNT) and (d) Armchair (a-PNT).

In particular, phosphorene is highly anisotropic, which manifests itself in interesting chirality-dependent properties. For instance, charge carriers in the zigzag direction are several times heavier than charge carriers in the armchair direction (see Figure 1b).[9, 24, 25] Consequently, the wrapping of phosphorene into phosphorene nanotubes

(PNTs) results in unique tunable electronic properties that are not present in conventional carbon nanotubes. In a recent study published this year, Guan *et al*[21]. claimed that most armchair and zigzag PNTs possess *indirect* bandgaps with a semiconducting to semi-metallic transition in the zigzag family of PNTs.[21] However, these results contradict previous studies by Guo *et al.*[19] and others[26] which asserted that all armchair and zigzag PNTs are semiconductors with *direct* bandgaps. To shed light and bring closure to these conflicting results, we present a thorough analysis of the electronic properties of these PNTs using extremely large-scale density functional theory (DFT) calculations in conjunction with large TZVP all-electron basis sets (up to 200 atoms and 4,400 orbitals). *In contrast to both of these previous studies*, we find that the electronic properties of these PNTs are surprisingly much more complex than either study claims. Specifically, we show that all PNTs exhibit *both* indirect and direct bandgaps at certain sizes, which is (1) further corroborated across several exchange-correlation functionals and (2) rationalized by a detailed comparison of analogous orbital interactions in the various PNTs and the parent phosphorene sheet.

All of the DFT calculations in this study were carried out with a massively parallelized version of the CRYSTAL14 program[27] with a large TZVP all-electron basis set,[28] which we have previously used in large-scale calculations of other massive nanotube structures.[29, 30] We have chosen to use the B3LYP hybrid functional[31] for all of the PNTs in this work since our own benchmark calculations on planar phosphorene give a direct bandgap of 2.25 eV (See Supporting Information), which is in excellent agreement with the 2.31 eV bandgap obtained from a recent many-body $G_0W_0$ (Green's function screened Coulomb) approach.[32] Most importantly, we also performed additional calculations with the range-separated HSE06[33] and the pure semilocal PBE[34] functionals to verify that our findings were not due to our specific choice of exchange-correlation potential, which further demonstrate that our results for all of the PNTs are completely general. It is also worth noting that all of the three different functionals surveyed in this study produced nearly identical geometries, with deviations of less than 1 Å in the geometry-optimized radius for even the largest (0,25) armchair nanotube (see the Supporting Information for plots comparing the nanotube radii obtained from the PBE, HSE06, and B3LYP functionals). Additional details of our calculations are given in the Computational Details section.

As shown in Figure 1, a single-walled PNT can be conceptualized as rolling up a phosphorene monolayer into a tube. Similarly to carbon nanotubes, each PNT can be completely specified by a pair of integer indices ($n,m$) that define a rolling vector $\mathbf{R} = n\mathbf{a_1}+m\mathbf{a_2}$, where $\mathbf{a_1}$, $\mathbf{a_2}$ are the lattice vectors of the monolayer. In this study, we consider two PNT chiralities, armchair ($0,m$) and zigzag ($n,0$), denoted by a-PNT and z-PNT, respectively. We note that for these chiralities, $m$ and $n$ denote the number of phosphorene units around the tube circumference.

Figure 2 shows the electronic band structure of selected PNTs along the irreducible Brillouin zone defined by the high-symmetry points Γ and X in momentum space (band structures for all PNTs and the various DFT methods can be found in the Supporting Information). In agreement with previous studies, a-PNTs consistently possess larger band gaps than z-PNTs, a trend that becomes more appreciable at smaller sizes. In stark contrast to the study by Guan *et al.*,[21] however, we observe that all PNTs are semiconducting. The only possible semi-metallic transition is predicted by the PBE functional, which is known to underestimate band gaps. For this reason, and because this transition would occur at extremely small nanotube diameters that may not be structurally stable, we conclude that all of the PNTs are semiconducting.

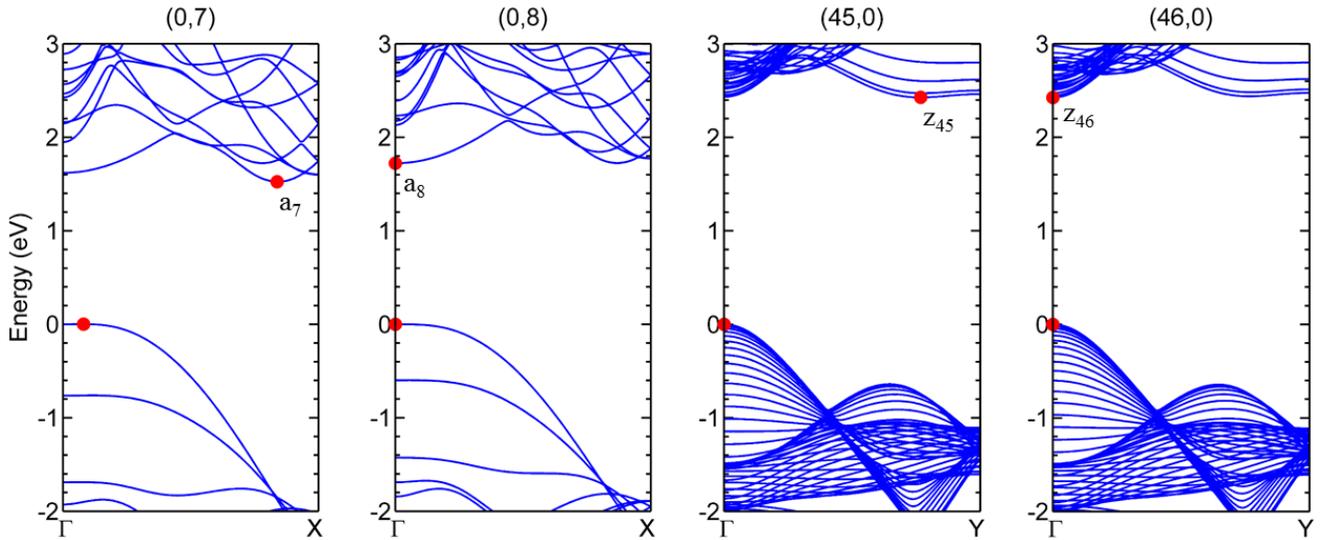

**Figure 2.** Electronic band structures of selected PNTs at the direct-to-indirect transition, with red dots indicating the VBM and CBM. The (0,8) a-PNT and (46,0) z-PNT possess direct band gaps, while the (0,7) a-PNT and (45,0) z-PNT possess indirect band gaps.

As expected, the band structures of large-diameter PNTs approach the infinite limit of the phosphorene sheet, with a direct bandgap at the Γ point. Interestingly, however, we find that below a certain size threshold ((0,8) and (46,0) for a-PNTs and z-PNTs, respectively) all of the PNTs exhibit an indirect bandgap, with a significant shift in the conduction band minimum (CBM) away from Γ. This dramatic change in electronic structure from a direct to indirect bandgap is in stark contrast to both of the previous studies by Guan[21] and Guo[19] and is further verified with the HSE06 and PBE functionals (See Supporting Information). In particular, we first observe that this direct-to-indirect transition occurs at a significantly larger diameter for z-PNTs than for a-PNTs. We attribute this effect to the considerably higher strain in z-PNTs (see Figure 3), which has been confirmed by past studies[19, 21, 35, 36] and by our own strain energy calculations (computed with the expression $E_{\text{nanotube}} - n \cdot E_{\text{sheet}}$, where $E_{\text{nanotube}}$ is the electronic energy of the optimized nanotube, $E_{\text{sheet}}$ is the electron energy of the phosphorene sheet, and $n$ is the number of repeat units along the PNT circumference; ). A natural question arises: What is the underlying electronic mechanism that causes this direct-to-indirect transition? In other words, why does increasing the strain in these tubes, *i.e.*, decreasing the tube diameter, shift the CBM (and VBM for a-PNTs) away from Γ?

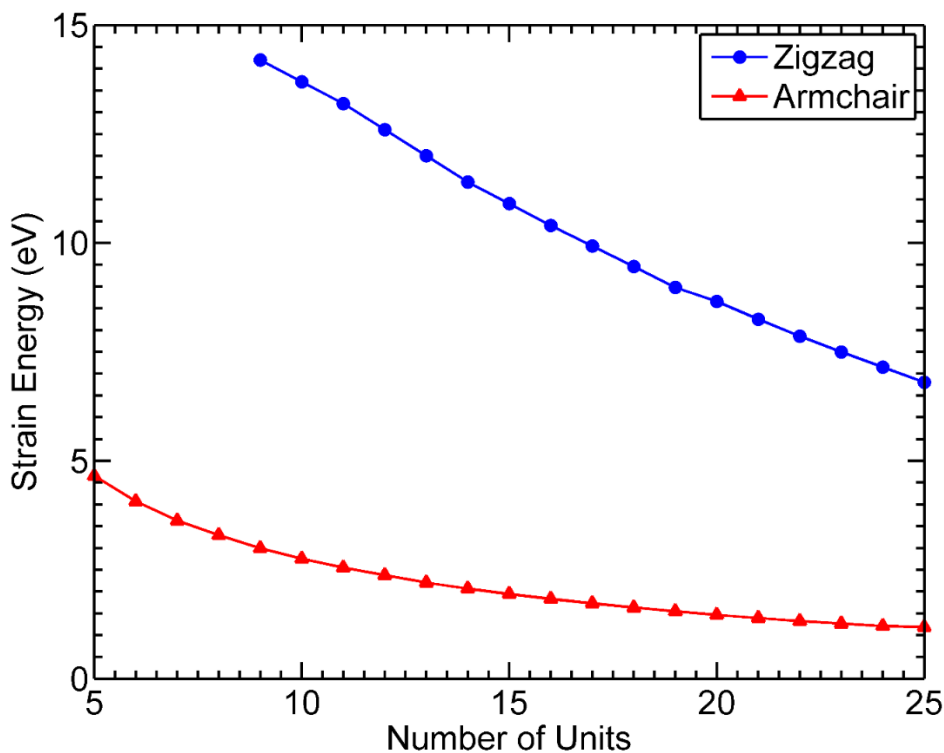

**Figure 3.** Plot of the strain energy as a function of the number of phosphorene units around the circumference (*i.e., m* or *n*).

To explore this question further and rationalize our results, we note that the wrapping of the phosphorene sheet into z-PNTs and a-PNTs is analogous to the stretching of the sheet in the *x* and *y* directions, respectively (cf. Figure 1). In a recent study, Peng *et al.* showed that stretching phosphorene in either direction causes bandgap transitions similar to the direct-to-indirect transition that we observe in the PNTs.[9] Thus, we associate the strain induced by stretching the sheet to the strain accompanied by decreasing the size of the nanotubes. As demonstrated by Peng *et al.* and by our own calculations (see Figure 4), stretching the sheet along both directions causes a shift in the CBM away from Γ. In particular, for each direction, *the CBM shifts in the symmetry direction of the corresponding nanotube;* namely, stretching the sheet in the zigzag direction causes the CBM to shift toward the Y point and stretching in the armchair direction causes the CBM to shift toward the X point.

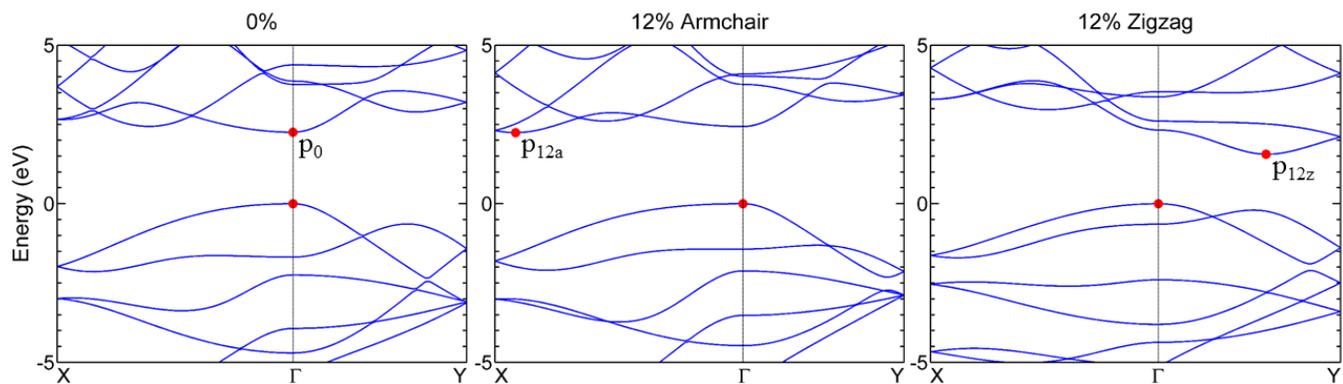

**Figure 4.** Band structure of phosphorene sheet: *(Left)* Unstretched, *(Center)* Stretched in the armchair direction (*y*), and *(Right)* Stretched in the zigzag direction (*x*).

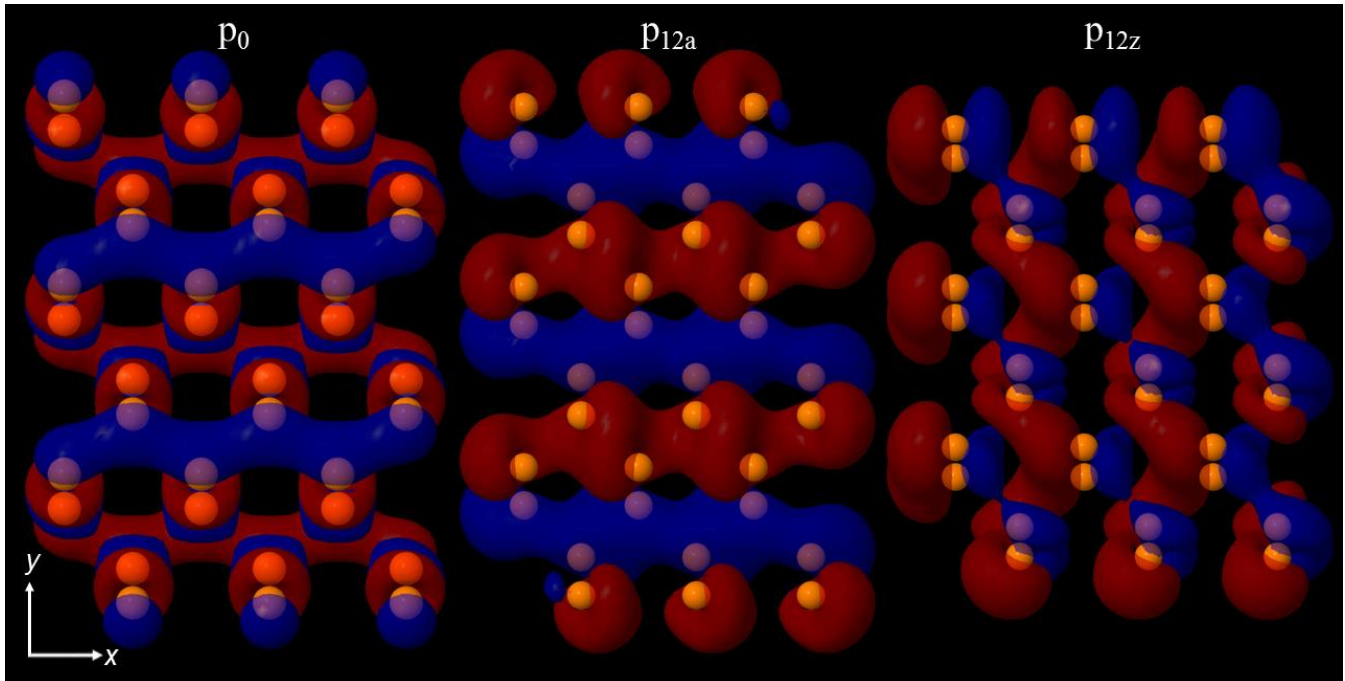

**Figure 5.** Bloch orbitals at the CBM of phosphorene sheet: *(Left)* Unstretched, *(Center)* Stretched in the armchair direction (*y*), *(Right)* Stretched in the zigzag direction (*x*).

This direct-to-indirect transition occurs because as the sheet is stretched, there is a competition in energetic stability between the orbital at the new CBM and the orbital at the old CBM (at $\Gamma$). When stretched, the overlap between orbitals in the stretching direction decreases, leading naturally to the destabilization (*i.e.,* increase in energy) of bonding orbitals and the stabilization (*i.e.,* decrease in energy) of antibonding orbitals. In Figure 5, we plot the orbitals corresponding to the CBM in the unstretched sheet at $\Gamma$ (denoted by $p_0$) and the CBM in the stretched sheets at the corresponding points (indicated by $p_{12a}$ and $p_{12z}$). From these plots, we see that stretching in the zigzag (*x*) direction destabilizes $p_0$ because the orbital is bonding in that direction, but stabilizes $p_{12z}$ because it is antibonding in that direction. Along the same lines, stretching in the armchair (*y*) direction does not affect $p_0$ significantly because it is nonbonding in that direction, but stabilizes $p_{12a}$ because it is antibonding in that direction.

Following a similar analysis, in Figure 6 we plot the CBM for the PNTs immediately before ($a_8$ and $z_{46}$) and after ($a_7$ and $z_{45}$) the direct-to-indirect transition. Before the transition, the strain in the PNTs is not large enough to significantly alter the band structure; hence, the CBM of these tubes resembles the CBM of the phosphorene sheet wrapped in the corresponding directions (see Figure 6). Below these sizes, however, the CBM shifts away from Γ, producing an indirect bandgap. In the a-PNTs, decreasing the nanotube size stabilizes both the orbitals at $a_7$ and $a_8$, but the orbital at $a_7$ stabilizes at a faster rate. As shown in Figure 5, $a_8$ is nonbonding around the circumference, while $a_7$ is antibonding around the circumference. Thus, increasing the strain has a much stronger stabilizing effect on $a_7$ than on $a_8$, leading to the shift in the CBM from $a_8$ to $a_7$ as the size decreases. For the (0,7) PNT, $a_7$ is stabilized and becomes the CBM. In the z-PNTs, decreasing the nanotube size destabilizes $z_{46}$ but stabilizes $z_{45}$. While both of these orbitals are bonding around the circumference, $z_{46}$ has stronger bonding character (*i.e.*, higher orbital overlap) in the outer ring of the PNT *and* in the inner ring of the PNT (see the blue and red Bloch orbitals in Figure 6). However, $z_{45}$ is bonding only along the outer ring. Consequently, increasing the strain has a stronger destabilizing effect on $z_{46}$, leading to the shift in the CBM to $z_{45}$. It is also interesting to note that unlike the CBM of the (0,7) a-PNT, the CBM of the (45,0) z-PNT does not resemble the CBM of the sheet stretched in the corresponding direction which we anticipate is due to the extremely high strain in the z-PNTs.

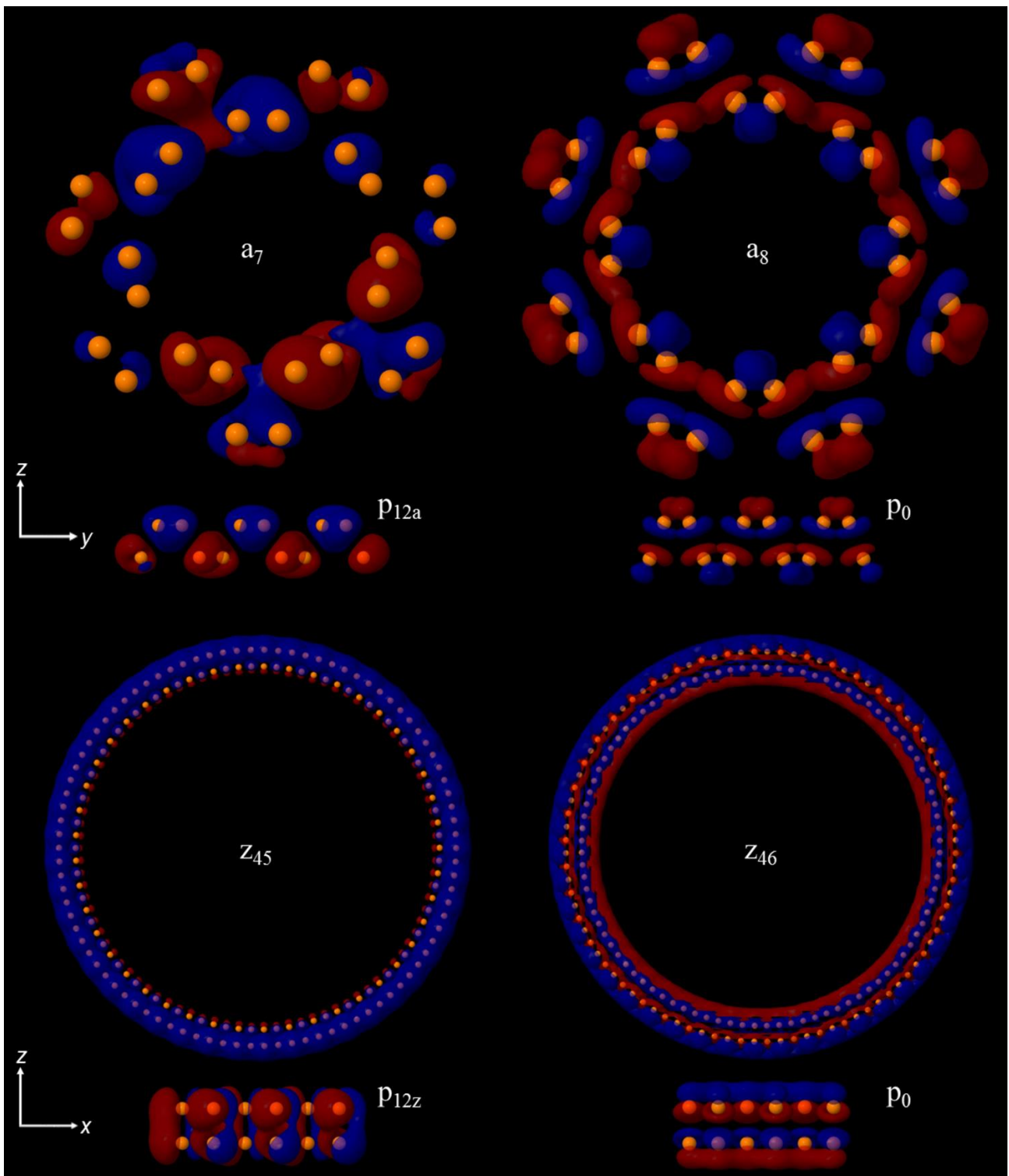

**Figure 6.** Bloch orbitals at the CBM of both the PNTs and the corresponding stretched sheets (shown below each tube). Prior to the direct-to-indirect transition, the CBM of the PNTs ($a_8$ and $z_{46}$) resembles the CBM of the unstretched sheet ($p_0$) viewed along the corresponding ($x$ or $y$) direction. After the transition, the CBM of the PNTs ($a_7$ and $z_{45}$) resembles the CBM of the sheet stretched in the corresponding direction ($p_{12a}$ and $p_{12z}$).

With these large-scale calculations and detailed orbital analyses, we have demonstrated that the electronic properties of phosphorene nanotubes are surprisingly rich and much more complex than previously claimed. Contrary to previous reports, we find that phosphorene nanotubes exhibit a direct-to-indirect bandgap transition as the nanotube diameter decreases – *a unique property that was not clearly identified in any of the prior studies.* This intricate transition arises from a competition in the energetic stability of different conduction band orbitals as the nanotube diameter decreases, which is further supported by detailed comparisons of analogous orbital interactions in the various PNTs and the parent phosphorene sheet. In particular, we have identified specific size regimes ((0,7) and (45,0) for a-PNTs and z-PNTs, respectively) where the CBM dramatically shifts away from the Γ point, resulting in an indirect bandgap. We also note that this direct-to-indirect transition is qualitatively different than the transition that has been shown to arise from axial strain in these PNTs.[19] The presence of both types of band gaps in PNTs is particularly noteworthy since these intrinsic electronic properties allow a wider range of tunability that is not present in conventional carbon nanotubes (which only possess direct band gaps). Specifically, our new calculations of *both* direct and indirect band gaps in PNTs have direct implications for applications that require either *(i)* a fast charge recombination and high light absorption (*i.e.*, a direct band gap) such as light-emitting and laser diodes, or *(ii)* a slow recombination and large diffusion length (*i.e.*, an indirect band gap) such as solar cells. We have shown that our findings are completely general, and extensive calculations across several exchange-correlation functionals with large all-electron basis sets verify our conclusions. Most importantly, our results and analyses resolve a long-standing question on the electronic properties of phosphorus nanotubes and brings closure to previously conflicting findings in these unique phosphorene nanostructures.

**Computational Details**

All calculations were carried out with a massively-parallelized version of the CRYSTAL14 program[37], which has the capability of using both all-electron Gaussian-type orbitals and exact Hartree-Fock exchange within periodic boundary conditions. The presence of nonlocal Hartree-Fock exchange is especially important for obtaining accurate electronic properties (in particular, band gaps) for periodic systems since the incorporation of Hartree-Fock exchange can partially correct for electron-delocalization errors inherent to both LDA (local density approximation) and GGA (generalized gradient approximation) exchange-correlation functionals.[29] To ensure that our calculations for the PNTs were completely general and not an artifact of the specific DFT method used, we evaluated a wide range of functionals that contain varying portions of Hartree-Fock exchange: (1) PBE, a non-empirical GGA functional that contains no exchange, (2) HSE06, which contains short-range exchange only, and (3) B3LYP, a popular global hybrid that contains a 20% fraction of Hartree-Fock exchange. Geometries for all of the phosphorene nanotubes were optimized using a large TZVP all-electron basis set[28] with one-dimensional periodic boundary conditions along the tube axis. At the optimized geometries, single-point calculations were performed with all three functionals with 100 $k$ points along the one-dimensional Brillouin zone to obtain the resulting electronic band structures.

**Supporting Information Available:** Structural and electronic properties of the parent phosphorene sheet, comparison of geometry-optimized radii obtained from the PBE, HSE06, and B3LYP functionals, and bandstructures for all the phosphorene nanotubes calculated with the various functionals.

## Acknowledgements

S.I.A. acknowledges support from a NASA MIRO Fields Fellowship. S.I.A. and B.M.W. acknowledge the National Science Foundation for the use of supercomputing resources through the Extreme Science and Engineering Discovery Environment (XSEDE), Project No. TG-ENG160024.

## References

(1)    Novoselov, K. S.; Geim, A. K.; Morozov, S. V.; Jiang, D.; Zhang, Y.; Dubonos, S. V.; Grigorieva, I. V.; Firsov, A. A. Electric Field Effect in Atomically Thin Carbon Films. *Science* **2004,** *306*, 666-669.
(2)    Liu, H.; Neal, A. T.; Zhu, Z.; Luo, Z.; Xu, X. F.; Tomanek, D.; Ye, P. D. Phosphorene: An Unexplored 2d Semiconductor with a High Hole Mobility. *ACS Nano* **2014,** *8*, 4033-4041.
(3)    Liang, L. B.; Wang, J.; Lin, W. Z.; Sumpter, B. G.; Meunier, V.; Pan, M. H. Electronic Bandgap and Edge Reconstruction in Phosphorene Materials. *Nano Lett.* **2014,** *14*, 6400-6406.
(4)    Tran, V.; Soklaski, R.; Liang, Y. F.; Yang, L. Layer-Controlled Band Gap and Anisotropic Excitons in Few-Layer Black Phosphorus. *Phys. Rev. B* **2014,** *89*, 235319.
(5)    Li, L. K.; Yu, Y. J.; Ye, G. J.; Ge, Q. Q.; Ou, X. D.; Wu, H.; Feng, D. L.; Chen, X. H.; Zhang, Y. B. Black Phosphorus Field-Effect Transistors. *Nat. Nanotechnol.* **2014,** *9*, 372-377.

(6) Koenig, S. P.; Doganov, R. A.; Schmidt, H.; Neto, A. H. C.; Ozyilmaz, B. Electric Field Effect in Ultrathin Black Phosphorus. *Appl. Phys. Lett.* **2014,** *104*, 103106.
(7) Das, S.; Zhang, W.; Demarteau, M.; Hoffmann, A.; Dubey, M.; Roelofs, A. Tunable Transport Gap in Phosphorene. *Nano Lett.* **2014,** *14*, 5733-5739.
(8) Wei, Q.; Peng, X. H. Superior Mechanical Flexibility of Phosphorene and Few-Layer Black Phosphorus. *Appl. Phys. Lett.* **2014,** *104*, 251915.
(9) Peng, X. H.; Wei, Q.; Copple, A. Strain-Engineered Direct-Indirect Band Gap Transition and Its Mechanism in Two-Dimensional Phosphorene. *Phys. Rev. B* **2014,** *90*, 085402.
(10) Ong, Z. Y.; Cai, Y. Q.; Zhang, G.; Zhang, Y. W. Strong Thermal Transport Anisotropy and Strain Modulation in Single-Layer Phosphorene. *J. Phys. Chem. C* **2014,** *118*, 25272-25277.
(11) Fei, R. X.; Yang, L. Lattice Vibrational Modes and Raman Scattering Spectra of Strained Phosphorene. *Appl. Phys. Lett.* **2014,** *105*, 083120.
(12) Kou, L. Z.; Frauenheim, T.; Chen, C. F. Phosphorene as a Superior Gas Sensor: Selective Adsorption and Distinct I-V Response. *J. Phys. Chem. Lett.* **2014,** *5*, 2675-2681.
(13) Ziletti, A.; Carvalho, A.; Campbell, D. K.; Coker, D. F.; Neto, A. H. C. Oxygen Defects in Phosphorene. *Phys. Rev. Lett.* **2015,** *114*, 046801.
(14) Kulish, V. V.; Malyi, O. I.; Persson, C.; Wu, P. Adsorption of Metal Adatoms on Single-Layer Phosphorene. *Phys. Chem. Chem. Phys.* **2015,** *17*, 992-1000.
(15) Dai, J.; Zeng, X. C. Bilayer Phosphorene: Effect of Stacking Order on Bandgap and Its Potential Applications in Thin-Film Solar Cells. *J. Phys. Chem. Lett.* **2014,** *5*, 1289-1293.
(16) Padilha, J. E.; Fazzio, A.; da Silva, A. J. R. Van der Waals Heterostructure of Phosphorene and Graphene: Tuning the Schottky Barrier and Doping by Electrostatic Gating. *Phys. Rev. Lett.* **2015,** *114*, 066803.
(17) Liu, Q. H.; Zhang, X. W.; Abdalla, L. B.; Fazzio, A.; Zunger, A. Switching a Normal Insulator into a Topological Insulator Via Electric Field with Application to Phosphorene. *Nano Lett.* **2015,** *15*, 1222-1228.
(18) Das, S.; Demarteau, M.; Roelofs, A. Ambipolar Phosphorene Field Effect Transistor. *ACS Nano* **2014,** *8*, 11730-11738.
(19) Guo, H. Y.; Lu, N.; Dai, J.; Wu, X. J.; Zeng, X. C. Phosphorene Nanoribbons, Phosphorus Nanotubes, and Van Der Waals Multilayers. *J. Phys. Chem. C* **2014,** *118*, 14051-14059.
(20) Wu, Q. Y.; Shen, L.; Yang, M.; Cai, Y. Q.; Huang, Z. G.; Feng, Y. P. Electronic and Transport Properties of Phosphorene Nanoribbons. *Phys. Rev. B* **2015,** *92*, 035436.
(21) Guan, L. X.; Chen, G. F.; Tao, J. G. Prediction of the Electronic Structure of Single-Walled Black Phosphorus Nanotubes. *Phys. Chem. Chem. Phys.* **2016,** *18*, 15177-15181.
(22) Tran, V.; Yang, L. Scaling Laws for the Band Gap and Optical Response of Phosphorene Nanoribbons. *Phys. Rev. B* **2014,** *89*, 245407.
(23) Han, X. Y.; Stewart, H. M.; Shevlin, S. A.; Catlow, C. R. A.; Guo, Z. X. Strain and Orientation Modulated Bandgaps and Effective Masses of Phosphorene Nanoribbons. *Nano Lett.* **2014,** *14*, 4607-4614.
(24) Qiao, J. S.; Kong, X. H.; Hu, Z. X.; Yang, F.; Ji, W. High-Mobility Transport Anisotropy and Linear Dichroism in Few-Layer Black Phosphorus. *Nat. Commun.* **2014,** *5*, 4475.
(25) Fei, R. X.; Yang, L. Strain-Engineering the Anisotropic Electrical Conductance of Few-Layer Black Phosphorus. *Nano Lett.* **2014,** *14*, 2884-2889.
(26) Yu, S.; Zhu, H.; Eshun, K.; Arab, A.; Badwan, A.; Li, Q. L. A Computational Study of the Electronic Properties of One-Dimensional Armchair Phosphorene Nanotubes. *J. Appl. Phys. (Melville, NY, U.S.)* **2015,** *118*, 164306.
(27) Dovesi, R.; Orlando, R.; Erba, A.; Zicovich-Wilson, C. M.; Civalleri, B.; Casassa, S.; Maschio, L.; Ferrabone, M.; De La Pierre, M.; D'Arco, P.*et al.* Crystal14: A Program for the Ab Initio Investigation of Crystalline Solids. *Int. J. Quantum Chem.* **2014,** *114*, 1287-1317.
(28) Peintinger, M. F.; Oliveira, D. V.; Bredow, T. Consistent Gaussian Basis Sets of Triple-Zeta Valence with Polarization Quality for Solid-State Calculations. *J. Comput. Chem.* **2013,** *34*, 451-459.
(29) Allec, S. I.; Ilawe, N. V.; Wong, B. M. Unusual Bandgap Oscillations in Template-Directed Π-Conjugated Porphyrin Nanotubes. *J. Phys. Chem. Lett.* **2016,** *7*, 2362-2367.

(30)	Pari, S.; Cuellar, A.; Wong, B. M. Structural and Electronic Properties of Graphdiyne Carbon Nanotubes from Large-Scale Dft Calculations. *J. Phys. Chem. C* **2016,** *120*, 18871-18877.
(31)	Becke, A. D. Density-Functional Thermochemistry .3. The Role of Exact Exchange. *J. Chem. Phys.* **1993,** *98*, 5648-5652.
(32)	Çakır, D.; Sevik, C.; Peeters, F. M. Significant Effect of Stacking on the Electronic and Optical Properties of Few-Layer Black Phosphorus. *Phys. Rev. B* **2015,** *92*, 165406.
(33)	Heyd, J.; Scuseria, G. E. Efficient Hybrid Density Functional Calculations in Solids: Assessment of the Heyd-Scuseria-Ernzerhof Screened Coulomb Hybrid Functional. *J. Chem. Phys.* **2004,** *121*, 1187-1192.
(34)	Perdew, J. P.; Burke, K.; Ernzerhof, M. Generalized Gradient Approximation Made Simple. *Phys. Rev. Lett.* **1996,** *77*, 3865-3868.
(35)	Cai, K.; Wan, J.; Wei, N.; Cai, H. F.; Qin, Q. H. Thermal Stability of a Free Nanotube from Single-Layer Black Phosphorus. *Nanotechnology* **2016,** *27*, 235703.
(36)	Liao, X. B.; Hao, F.; Xiao, H.; Chen, X. Effects of Intrinsic Strain on the Structural Stability and Mechanical Properties of Phosphorene Nanotubes. *Nanotechnology* **2016,** *27*, 215701.
(37)	Lu, X.; Chen, Z. F. Curved Pi-Conjugation, Aromaticity, and the Related Chemistry of Small Fullerenes (< C-60) and Single-Walled Carbon Nanotubes. *Chem. Rev. (Washington, DC, U.S.)* **2005,** *105*, 3643-3696.